\documentclass[prl,twocolumn,superscriptaddress]{revtex4}
\usepackage{epsfig}
\usepackage{times}
\usepackage{amsmath}
\usepackage{amsfonts}
\usepackage{amssymb}
\usepackage[usenames]{color}
\usepackage{graphicx}

\newcommand{\beq}{\begin{equation}}
\newcommand{\eeq}{\end{equation}}
\newcommand{\beqa}{\begin{eqnarray}}
\newcommand{\eeqa}{\end{eqnarray}}

\begin{document}
\title{Magnet field sensing beyond the standard quantum limit
 under the effect of decoherence}
\author{Yuichiro Matsuzaki}
\affiliation{Department of Materials, University of Oxford, OX1 3PH,
U. K.}
\author{Simon C. Benjamin\footnote{s.benjamin@qubit.org}}
\affiliation{Department of Materials, University of Oxford, OX1 3PH, U. K.}
\affiliation{Centre for Quantum Technologies, National University of Singapore, 3 Science Drive 2, Singapore 117543.}
\author{Joseph Fitzsimons}
\affiliation{Centre for Quantum Technologies, National University of Singapore, 3 Science Drive 2, Singapore 117543.}

\begin{abstract}
   Entangled states can potentially be used to outperform the standard quantum limit which every classical sensor is bounded by.  However, entangled states are very susceptible to decoherence, and so it is not clear whether one can really create a superior sensor to classical technology via a quantum strategy which is subject to the effect of realistic noise. This paper presents an investigation of how a quantum sensor composed of many spins is affected by independent dephasing. We adopt general noise models including non-Markovian effects, and in these noise models the performance of the sensor depends crucially on the exposure time of the sensor to the field. We have found that, by choosing an appropriate exposure time within non-Markovian time region, an entangled sensor does actually beat the standard quantum limit. Since independent dephasing is one of the most typical sources of noise in many systems, our results suggest a practical and scalable approach to beating the standard quantum limit.
\end{abstract}

\maketitle

Entanglement has proven itself to be one of the most intriguing aspects
of quantum mechanics, and its study has lead to profound advances in our
understanding of physics. Aside from these conceptual advances, the
exploitation of entanglement has lead to a number of technical advances
both in computation and communication \cite{nielsen00}, and more
recently in metrology \cite{giovannetti2004quantum}. In quantum metrology, entanglement has been used to demonstrate enhanced accuracy both in detecting the phase induced by unknown optical elements and for accurately measuring an unknown magnetic field. It is this latter case which is the focus of the present paper, and so we will adopt the terminology of field sensing.

 In order to estimate an unknown field, one usually prepares a probe
 system composed of $L$ distinct local subsystems, exposes this to the field for a certain
 time, and measures the probe. Comparing the input with the output of the probe gives us an estimate of the field. Importantly, there is an uncertainty in the estimation, and this
 uncertainty is related to how the probe system is prepared.
 When the probe system is prepared in a separable state, the uncertainty decreases as
 $1/\sqrt{L}$ \cite{pezze2009entanglement,itano1993quantum} by the central limit theorem, a scaling known as the standard quantum limit. On the other hand, by
 preparing a highly entangled state, it is in principal possible to achieve an uncertainty that scales as $1/L$, known the Heisenberg limit \cite{giovannetti2004quantum,giovannetti2006quantumprlversion}.

 Most of the literature on quantum sensing focuses on using photons to
 probe an unknown optical element.
 Using a NOON state \cite{nagata2007beatingsimple,afek2010high,kok2002creation},
 it is possible to measure the unknown phase
 shift with higher resolution than the standard quantum limit. An $L$-photon NOON state can achieve a phase $L$ times as large as than in the case of a single photon over the same channel. 
 Recent publications
 \cite{leibfried2005creationsimple,jones2009magneticsimple,leibfried2004toward,schaffry2010ensemblebasedmetrology}
 have considered an analogous technique for field
 sensing with spins.
 In this paper, we consider an experiment involving a probe consisting of spin-$\frac{1}{2}$ systems.
The spin qubits can couple to the magnetic field and therefore one can estimate the value of the magnetic field by the probe. In order to obtain higher resolution than the standard quantum limit,
 one can use a GHZ state, as has been demonstrated in recent experiments \cite{leibfried2005creationsimple,jones2009magneticsimple,simmons2009magnetic0907,leibfried2004toward}.

 In solid state systems, one of the main barriers to realising such
 sensors is decoherence, which degrades the quantum coherence of the entangled states. GHZ states, in particular, are very
 susceptible to decoherence, and decohere more rapidly as the size of the state
 increases \cite{dur2004stabilitymacro,shajicaves2007qubit}. Therefore,
 it is not clear whether a quantum strategy can really outperform an
 optimal classical strategy under the effects of a realistic noise
 source. The effect of unknown but static field variations over the L spins has been studied by Jones {{\it et al}} \cite{jones2009magneticsimple}.  The uncertainty of the estimation depends on the exposure time of entangled states to the field, where they are affected by noise, and they have found that, for an optimal exposure time in their model, the scaling of the estimated value is $L^{-\frac{3}{4}}$ which beats the standard quantum limit.
However, the underlying assumption that the fields are static could be unrealistic for many systems, as actual noise in the laboratory may fluctuate with time.
 Huelga {\it{et al}} have included such temporal fluctuations of the field
 in their noise model \cite{huelga1997improvement} and have shown that
 GHZ states cannot beat the standard quantum limit under the effect of independent dephasing by adopting a Lindblad type master equation \cite{huelga1997improvement}. Even for the optimal exposure time, it was shown that the measurement uncertainty of a quantum strategy has the same scaling behavior as the standard quantum limit in their noise model. Since independent dephasing is the dominant error sources in many systems, these results seem to show that, practically, it would be impossible to beat the standard quantum limit with a quantum strategy.
 
 However, the model adopted by Huelga {\it{et al}} is a Markovian master
 equation \cite{GZ01b} which is valid in limited circumstances. The Markovian assumption will be violated when the correlation time of the noise is longer than the characteristic time of the system. For example, although a Markovian master equation predicts an exponential
 decay behavior, it is known that unstable systems show a quadratic
 decay in the time region shorter than a correlation time of the noise \cite{NakazatoNamikiPascazio01a,Schulman01a}.
 In this paper, we adopt independent dephasing models which include non-Markovian
 effects and we investigate how the uncertainty of the estimation is affected by such noise. We have found that, if the exposure time of the entangled state is within
 the non-Markovian region, a quantum strategy can indeed provide a scaling advantage over the optimal classical strategy.
 
 Let us summarize a quantum strategy to obtain the Heisenberg limit in an
 ideal situation without decoherence. A state prepared in
 $|+\rangle
 =\frac{1}{\sqrt{2}}(|0\rangle +|1\rangle )$ will have a phase factor in
 its non-diagonal term
 through being exposed in a magnetic field and so we have $\frac{1}{\sqrt{2}}(|0\rangle
 + e^{-i t\delta }|1\rangle ) $ where $\delta $ denotes the detuning
 between the magnetic field and the atomic transition.
 On the other hand, when one prepares a GHZ state
 $|\psi \rangle _{\text{GHZ}}=\frac{1}{\sqrt{2}}(|00\cdots 0\rangle +|11\cdots
     1\rangle )$
     and exposes this state to the field for a time $t$,
     the phase factor
     is amplified linearly as the
 size of the state increases as 
     \begin{eqnarray}
      |\psi (t)\rangle =\frac{1}{\sqrt{2}}(|00\cdots 0\rangle
       +e^{-iLt\delta }|11\cdots
       1\rangle ).
     \end{eqnarray}
     Therefore, the probability of finding the initial GHZ state after a time $t$ is given
by $ P=\frac{1}{2}+\frac{1}{2} \cos  (Lt\delta )$. In practice, one may
use control-not operations to map the accumulated phase to a single spin
for a convenient measurement \cite{jones2009magneticsimple}.
The variance of the estimated value is then given by
\begin{eqnarray}
 \Delta ^2\delta =\frac{P(1-P)/N}{|dP/d\delta |^2}\label{formula-uncertainty}
\end{eqnarray}
where $N$ is the number of experiments performed in this setting \cite{huelga1997improvement}.
For a given fixed time $T$, one can perform this experiment $T/t$ times
where $t$ is the exposure time, and so we have $N=T/t$.
Hence we obtain $|\Delta ^2\delta|=\frac{1}{TL^2t}$ and so the uncertainty in $\delta$ scales as $L^{-1}$, the Heisenberg limit. 

First let us consider the decoherence of a single qubit. Later, we will generalize to GHZ states. Our noise model represents random classical fields to induce dephasing. We consider an interaction Hamiltonian to denote a coupling with an environment such as 
 \begin{eqnarray}
  H_I=\lambda f(t)\hat{\sigma }_z \label{eq:1qubitH}
 \end{eqnarray}
where $f(t)$ is classical normalized Gaussian noise,
$\hat{\sigma }_z$ is a Pauli
operator of the system, and $\lambda $ denotes a coupling constant.
Also, we assume symmetric noise to satisfy $\overline{f(t)}=0$
where this over-line denotes the average over the ensemble of
the noise.
When we solve the Schr{\"o}dinger
equation in an interaction picture, we
obtain the following standard form 
\begin{eqnarray}
      &&\rho_I(t) -\rho _0
         =\sum_{n=1}^{\infty
   }(-i\lambda )^n\int_{0}^{t}dt_1\int_{0}^{t_1}dt_2\cdots
   \int_{0}^{t_{n-1}}dt_n\nonumber \\
   &&[H_I(t_1),[H_I(t_2),\cdots ,[H_I(t_n),\rho_0
   ]\cdots ]]\label{1-analytical} 
      \end{eqnarray}
where $\rho _0=|\psi \rangle \langle \psi |$ is an initial state and $\rho _I(t)$ is a state in the interaction picture. Throughout this paper, we restrict ourselves to the case where the system Hamiltonian commutes with the operator of noise, as this constitutes purely dephasing noise. By taking the average over the ensemble of the noise, we obtain
     \begin{eqnarray}
       &&\rho_I(t) -\rho _0
        =\sum_{n=1}^{\infty
  }\frac{1}{n!}(-i\lambda )^n\int_{0}^{t}dt_1\int_{0}^{t}dt_2\cdots
  \int_{0}^{t}dt_n\nonumber \\
  &&\overline{f(t_1)f(t_2)\cdots f(t_n)}[\hat{\sigma }_z,[\hat{\sigma
   }_z,\cdots ,[\hat{\sigma }_z,\rho_0
  ]\cdots ]]_n\label{2-analytical} 
     \end{eqnarray}
     where $[\hat{\sigma }_z,[\hat{\sigma
   }_z,\cdots ,[\hat{\sigma }_z,\rho_0
  ]\cdots ]]_n$ denotes the $n$-folded commutator of $\rho _0$ with
   $\hat{\sigma }_z$.
Since all higher order cumulants than
the second order are zero for Gaussian noise, $\overline{f(t_1)f(t_2)\cdots
f(t_n)}$ can be represented 
by a product of correlation functions \cite{Meeron01a}. Therefore, the decoherence caused by
Gaussian noise is characterized by a correlation function of the noise. For Markovian white noise, a correlation function becomes a delta function while a correlation function becomes a constant
for non-Markovian noise with an infinite correlation time such as $1/f$ noise. To include both noise models as special cases, we assume the correlation function
\begin{eqnarray}
 \overline{f(t_1)f(t_2)}=\frac{2}{\sqrt{\pi
}}e^{-\frac{|t_1-t_2|^2}{\tau _c^2}},
\end{eqnarray}
 where $\tau _c$ denotes the correlation time of the noise. In the limit $\tau _c \rightarrow 0$, this correlation function becomes a delta function, while in the limit of $\tau_c\rightarrow \infty$ it becomes constant.

Since $\overline{f(t_1)f(t_2)\cdots
f(t_n)}$ can be represented by a product of correlation functions,
we obtain 
          \begin{eqnarray}
       \rho_I(t) 
            &=&\sum_{n=0}^{\infty
  }\frac{(-\frac{1}{4} t\gamma (t) )^{n}}{n!}[\hat{\sigma
  }_z,[\hat{\sigma }_z,\cdots ,[\hat{\sigma }_z,\rho_0
  ]\cdots ]]_{2n}\nonumber \\
      &=&\sum_{s,s'=\pm 1}e^{-\frac{1}{4}|s-s'|^2\gamma (t) t} |s \rangle \langle s|\rho
  _0|s'\rangle \langle s'|\label{3-analytical}
     \end{eqnarray}
     where $|s\rangle $ is an eigenvector of $\hat{\sigma }_z$.
     Also, $\gamma (t)$ denotes the single qubit decoherence rate defined as
     \begin{eqnarray}
      \gamma (t)=\frac{4\lambda ^2\tau _c^2(-1+e^{-\frac{t^2}{\tau _c^2}})}{\sqrt{\pi
     }t} +4\lambda ^2\tau _c\text{erf}(\frac{t}{\tau _c})\label{decoherence-rate-definition}
     \end{eqnarray}
     where $\text{erf}(x)$ is the error function.
     Note that, for $t\gg \tau _c$, the decoherence
     rate becomes constant as
     $\gamma
     (t)\simeq 4\lambda ^2 \tau _c$. So, in this regime, we can derive a
     Markovian master equation from (\ref{3-analytical}), which has the
     same form as adopted by Huelga {\it{et al}} \cite{huelga1997improvement}.
     \begin{eqnarray}
      \frac{d\rho _I(t)}{dt}\simeq -\lambda ^2\tau _c[\hat{\sigma
       }_z[\hat{\sigma }_z,\rho _I(t)]]\ \ (t\gg \tau _c)
     \end{eqnarray}
     Note that, although our model can be approximated by Markovian noise in the long time limit ($t\gg \tau _c$), we are interested in the time periods $t\sim \tau _c$  and  $t\ll \tau _c$ where non-Markovian effects become relevant.

For an initial state $|+\rangle =\frac{1}{\sqrt{2}}(|0\rangle +|1\rangle
)$, the non-diagonal terms of the density matrix show a decay behavior of
$\langle 0|\rho _I(t)|1\rangle =\frac{1}{2}\text{exp}[-\gamma (t)t]$.
 \begin{figure}[h]
   \begin{center}
       \includegraphics[width=7.0cm]{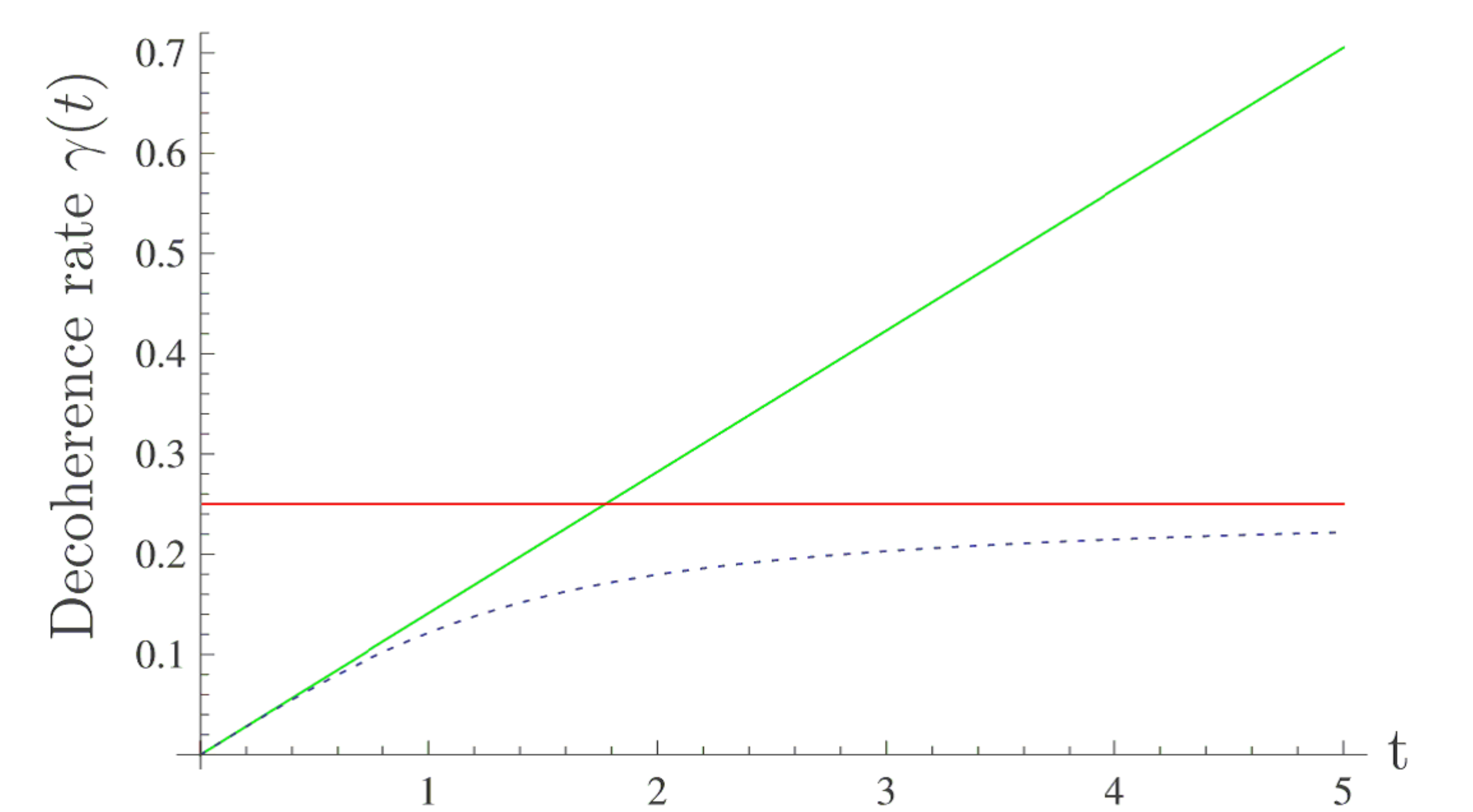}
   \end{center}
  \caption{
We have plotted the behavior of a single qubit decoherence rate for 
  $1/f$ noise
   $\frac{4}{\sqrt{\pi}}\lambda ^2  t$ (the green line),
  Markovian noise $4\lambda ^2
  \tau _c $ (the red line), and
  our classical noise
  model (blue dotted line) $\gamma(t)$ defined in (\ref{decoherence-rate-definition}) where $\lambda $ and $\tau_c$
  denote a coupling constant with the environment and a correlation time
  of the noise, respectively. For an $L$-qubit GHZ state, the decoherence
  rate becomes $L$ times greater than the value plotted here.
  Note that our noise
  model shows a transition from a non-Markovian quadratic decay to an
   exponential Markovian decay.
  Here, we fixed parameters as $\lambda =0.25$ and $\tau _c=1$.
  } \label{decay-behavior}
   \end{figure}
For $t\gg \tau _c$, 
the state shows an exponential decay $ \langle 0|\rho _I(t)|1\rangle
\simeq  \frac{1}{2}e^{-4\lambda ^2 \tau _c t}$. On the other hand,
we have $ \langle 0|\rho _I(t)|1\rangle \simeq
\frac{1}{2}e^{-\frac{4}{\sqrt{\pi}}\lambda ^2  t^2}$ (quadratic decay
behavior) for $t\ll \tau _c$ ,
which is the typical
decay behavior of $1/f$
noise \cite{YoshiharaHarrabiNiskanenNakamura01a,KakuyanagiMenoSaitoNakanoSembaTakayanagiDeppeShnirman01a,matsuzaki2010quantumzenontt}.
The behavior of the decoherence rate is illustrated in
Fig.~\ref{decay-behavior}.

We next consider decoherence of a GHZ state induced by random classical fields. Extending the Hamiltonian in (\ref{eq:1qubitH}), the interaction Hamiltonian denoting random classical fields for a many-qubit system is as follows:
\begin{eqnarray}
 H_I=\lambda \sum_{l=1}^{L}f_l(t)\hat{\sigma }_z(l)
\end{eqnarray}
where $f_l(t)$ denotes the noise acting at site $l$
and has the same characteristics as $f(t)$ mentioned above. Also, since
we consider independent noise, we have $\overline
{f_l(t)f_{l'}(t')}\propto \delta _{l,l'}$.
When we let a GHZ state be exposed to a magnetic field under the effect
of the random magnetic fields, the state will remain in
the subspace spanned by $\bigotimes ^L_{l=1}|0\rangle _l $ and $\bigotimes
^L_{l=1}|1\rangle _l$ because we are only considering phase noise.
Thus, we can use the same analysis as for a single qubit.
In the the Schr{\"o}dinger picture, we will obtain
\begin{eqnarray}
&& \rho (t)
  =\frac{1}{2}\big{(}\bigotimes ^L_{l=1}
   |0\rangle _l\langle 0|\big{)}
 +\frac{1}{2}e^{ iLt\delta - L\gamma (t)t}\big{(}\bigotimes ^L_{l=1}|0\rangle
 _l\langle 1|\big{)}\nonumber \\
  &+&\frac{1}{2}e^{ - iLt\delta -L\gamma (t)t}\big{(}\bigotimes ^L_{l=1}|1\rangle _l\langle 0|\big{)}
  +\frac{1}{2}\big{(}\bigotimes ^L_{l=1}|1\rangle _l\langle 1|\big{)}
\end{eqnarray}
where $\delta $ is the phase induced by the magnetic field to be measured.
It is worth mentioning that the decoherence rate for a $L$-qubit GHZ state
becomes $L$ times the decoherence rate $\gamma (t)$ of a single qubit.
Therefore, the variance of the estimated value becomes
\begin{eqnarray}
 \Delta ^2\delta
  &=& \frac{e^{2 \gamma (t)Lt}-1}{TL^2t\sin ^2(L t
  \delta)}+\frac{1}{TL^2t} 
\end{eqnarray}
where we use (\ref{formula-uncertainty}), and therefore
we obtain the following inequality by using $x \geq \sin x\geq \frac{2
}{\pi }x$ for $0\leq x\leq \frac{\pi }{2}$:
\begin{eqnarray}
 \frac{e^{2 \gamma (t)Lt}-1}{
  TL^4t^3 \delta ^2}
\leq  (\Delta ^2 \delta) - \frac{1}{TL^2t}
\leq \frac{\pi ^2(e^{2 \gamma (t)Lt}-1)}{
  4TL^4t^3\delta ^2}
  \label{triangle-omega}
\end{eqnarray}

These inequalities depend on both the system size $L$ and the choice of
exposure time $t$.
We wish to see if there is any choice of $t$ for which we beat the
standard quantum limit.
We find that this limit can indeed be beaten provided that we chose
shorter $t$ values for larger systems.
For example, suppose that we chose $t$ according to the rule
$t=sL^{-z}$, where $s$ is a constant with the dimension of time,
and $z$ is a non-negative real number whose optimal value we will
determine. Then, from (\ref{decoherence-rate-definition}), we have
\begin{eqnarray}
 |\gamma (t)-\frac{4s\lambda ^2}{\sqrt{\pi }L^{z}}|\leq
  \frac{12s^2\lambda ^2/\tau _c}{\sqrt{\pi }L^{z}(L^{2z}-s^2/\tau _c^2)}\label{gamma-scale-triangle}
\end{eqnarray}
for large $L$
and hence the decoherence rate scales as $\gamma (t)=\Theta(L^{-z})$. Throughout this paper, for a function $f(L)$, we say $f(L)=\Theta (L^n)$ if there exists positive constants $J$ and $K$ such that $JL^n \leq f(L)\leq KL^n$ is satisfied for all $L$.

In (\ref{triangle-omega}), the term in the exponential $\gamma
(t)tL$ goes to infinity as $L\rightarrow \infty $ for $z<1/2$, and so the
uncertainty $|\Delta ^2\delta |$ diverges, which means that
a large GHZ state becomes useless to estimate $\delta $. Therefore, we consider
  the case of $z\geq 1/2$.
By performing a Taylor expansion of $e^{2 \gamma (t)Lt}$, we obtain
\begin{eqnarray}
 \frac{e^{2 \gamma (t)Lt}-1}{
  L^4t^3}=\sum_{n=1}^{\infty }\frac{(2
  \gamma (t)Lt)^n}{L^4t^3\cdot n!}=\Theta(L^{z-3})\ \ \ \ \ \label{scaling-part-omega}
\end{eqnarray}
for $z\geq 1/2$
where we use (\ref{gamma-scale-triangle}).

So, from (\ref{triangle-omega}) and (\ref{scaling-part-omega}), we
obtain
\begin{eqnarray}
 \Delta ^2 \delta =\Theta(L^{-2+z})
\end{eqnarray}
for $z\geq \frac{1}{2}$. 
Therefore, when $z=\frac{1}{2}$, we achieve a scaling of the uncertainty
$|\Delta \delta |=\Theta(L^{-\frac{3}{4}})$ 
and this actually beats the standard quantum limit. Note that this is the same scaling
as the magnetic sensor under the effect of unknown static fields studied
in \cite{jones2009magneticsimple}, and so our result for the
fluctuating noise with time becomes a natural generalization of
their work.

The decoherence model described above is a classical one.
Next, we make use of a quantized model where the environment is modeled as a
continuum of field modes. 
The Hamiltonian of the system and the environment are defined as
\begin{equation}
 H=\frac{\delta }{2} \sum_{l=1}^{L}\hat{\sigma }_z(l)
  +\sum_{l,{{k}}}\omega _{{{k}}}
  \hat{b}^{\dagger }_{l,{{k}}}\hat{b}_{l,{{k}}}
  +\sum_{l,{{k}}}g _{{{k}}}\hat{\sigma }_z(l)
  \hat{b}^{\dagger }_{l,{{k}}}\hat{b}_{l,{{k}}}
\end{equation}
where $\hat{b}_{l,{{k}}}$ and $\hat{b}^{\dagger }_{l,{{k}}}$
denote annihilation and creation operators for the bosonic
field at a site $l$.
Also, since we consider independent noise, we assume that
$\hat{b}_{l,{{k}}}$  commutes with
$\hat{b}^{\dagger }_{l',{{k}}}$ for $l\neq l'$.
This model has been solved analytically by Palma {\it{et al}} \cite{PSE}
and the time evolution
of a GHZ state is given by
\begin{eqnarray}
 &&\rho (t)
  =
  \frac{1}{2}\big{(}\bigotimes ^L_{l=1}
   |0\rangle _l\langle 0|\big{)}  +\frac{1}{2}e^{iLt\delta -L\Gamma (t)t}\big{(}\bigotimes ^L_{l=1}
   |0\rangle _l\langle 1|\big{)}
  \nonumber \\
  &+&\frac{1}{2} e^{iLt\delta -L\Gamma (t)t}\big{(}\bigotimes ^L_{l=1}
   |1\rangle _l\langle 0|\big{)}
  +\frac{1}{2} \big{(}\bigotimes ^L_{l=1}
   |1\rangle _l\langle 1|\big{)}, \ \ \ \ \ \ \ \  
\end{eqnarray}
where 
$\Gamma (t)$ denotes a decoherence rate defined as
$\Gamma (t)=\frac{1}{t}\log (1+\omega _c^2t^2)+\frac{2}{t}\log
(\frac{\sinh (\pi Tt)}{\pi Tt})$ where we have taken the Boltzmann
constant $k_B=1$. Here, $T$ and $\omega _c$ denote the temperature and cut off frequency respectively. As we take $k_B=1$, the temperature $T$
has the same dimension as the frequency $\omega _c$.
We have a constant decoherence rate $\Gamma
(t)\simeq 2\pi T$ for $t\gg T^{-1}$, which signals Markovian
exponential decay, while we have $\Gamma (t)\simeq \omega _c^2t$
for $t\ll \omega _c^{-1}$, which is the characteristic decay of $1/f$ noise. So, by taking these limits, this model can
also encompass both Markovian noise and $1/f$ noise.
Given the calculations in the previous section,
the variance of the estimated value for the magnetic field becomes
\begin{eqnarray}
 \Delta ^2\delta  =\frac{(1+\omega _c^2t^2)^{2L}(\frac{\sinh (\pi Tt)}{\pi Tt})^{4L}-1}{
  TL^2t\cdot \sin ^2(L t \delta)}+\frac{1}{TL^2t}\ .
\end{eqnarray}
By performing a calculation exactly analogous to the case of the random classical field considered above, one can show that
the scaling law for the uncertainty becomes
$\Delta ^2\delta =\Theta(L^{-2+z}) $
 for $z\geq \frac{1}{2}$ when we take an exposure time $t=s/L^{z}$.
 On the other hand, the uncertainty will diverge
as $L$ increases for $z<\frac{1}{2}$.
Therefore, by taking $z=\frac{1}{2}$, the uncertainty scales as
$\Theta(L^{-\frac{3}{4}})$ and so one can again beat the standard quantum
limit as before.

We now provide an intuitive reason why
the uncertainty of the estimation diverges for large $L$ when
$z$ is below $\frac{1}{2}$ in both of noise models.
It has been shown that an unstable state always shows a quadratic decay
behavior in an initial time region \cite{NakazatoNamikiPascazio01a,Schulman01a},
and therefore the scaling behavior of the fidelity of a single qubit
should be $F =\langle \psi |\rho
(t)|\psi \rangle =1-Ct^2 +O(t^3)$
where $C$ is a constant.
So the scaling behavior of the fidelity becomes
$F=1-CLt^2
+O(t^3)$ for multipartite entangled states
under the effect of independent
noise \cite{duan1997perturbativefidelity}. If we take a time as $t=sL^{-z}$, to first order we obtain an infidelity of $1-F\simeq Cs^2L^{1-2z}$. So this
infidelity becomes larger
as $L$ increases for $z<\frac{1}{2}$, which means coherence of this state will be almost
completely destroyed for a large GHZ state. On the other hand, as long as we have $z\geq
\frac{1}{2}$, the infidelity can be bounded by a constant even for a
large $L$ and so the coherence of the state
will be preserved, which can be utilized for a quantum magnetic sensor.

Finally, we remark on the prospects for experimental realization of our model.
To experimentally realise such a sensor, one has to generate a GHZ state, expose the state in a magnetic
field, and measure the state, before the state shows an exponential decay.
Although it has been shown that an unstable system shows a
quadratic decay behavior in the initial time period shorter than a correlation
time of the noise \cite{NakazatoNamikiPascazio01a,Schulman01a}, it is
difficult to observe such quadratic decay behavior
experimentally, because the correlation time of the noise
is usually much shorter than the typical time resolution of a measurement
apparatus for current technology.
After showing the quadratic decay behavior, unstable systems usually show an
exponential decay \cite{NakazatoNamikiPascazio01a}
and, in this exponential decay region, it is not possible to beat the
standard quantum limit \cite{huelga1997improvement}.
However, it is known that $1/f$ noise has an infinite
correlation time and one doesn't observe an exponential decay of a system
affected by $1/f$ noise \cite{matsuzaki2010quantumzenontt}.
Therefore, a system dominated by such noise
would be
suitable for the first experimental demonstration of our model.
For example, it is known that
nuclear spins of donor atoms in doped silicon devices, which have been
proposed as qubits for quantum computation \cite{K01a}, are dephased
mainly by
1/f noise
\cite{K01a,ladd2005coherence} and so they may prove suitable to
demonstrate our prediction.

In conclusion, we have shown that, under the effect of independent dephasing, one can obtain a magnetic sensor whose
uncertainty scales as $\Theta(L^{-\frac{3}{4}})$ and therefore beats the
standard quantum limit of $L^{-\frac{1}{2}}$. We determine that, to outperform
a classical strategy, the exposure time of the entangled states to the
field should be within the non-Markovian time region where the decoherence
behavior doesn't show exponential decay.
Since the noise models adopted here are quite general, our
results suggest a scalable method to beat the
standard quantum limit in a realistic setting.

The authors thank M. Schaffry and E. Gauger for useful discussions.
This research is supported by the National Research Foundation and Ministry of Education, Singapore. YM is supported by the Japanese Ministry of Education, Culture, Sports, Science and 
 Technology. 

\begin{thebibliography}{27}
\expandafter\ifx\csname natexlab\endcsname\relax\def\natexlab#1{#1}\fi
\expandafter\ifx\csname bibnamefont\endcsname\relax
  \def\bibnamefont#1{#1}\fi
\expandafter\ifx\csname bibfnamefont\endcsname\relax
  \def\bibfnamefont#1{#1}\fi
\expandafter\ifx\csname citenamefont\endcsname\relax
  \def\citenamefont#1{#1}\fi
\expandafter\ifx\csname url\endcsname\relax
  \def\url#1{\texttt{#1}}\fi
\expandafter\ifx\csname urlprefix\endcsname\relax\def\urlprefix{URL }\fi
\providecommand{\bibinfo}[2]{#2}
\providecommand{\eprint}[2][]{\url{#2}}

\bibitem[{\citenamefont{Nielsen and Chuang}(2000)}]{nielsen00}
\bibinfo{author}{\bibfnamefont{M.~A.} \bibnamefont{Nielsen}} \bibnamefont{and}
  \bibinfo{author}{\bibfnamefont{I.~L.} \bibnamefont{Chuang}},
  \emph{\bibinfo{title}{Quantum Computation and Quantum Information}}
  (\bibinfo{publisher}{{Cambridge University Press}}, \bibinfo{year}{2000}),
  ISBN \bibinfo{isbn}{521635039}.

\bibitem[{\citenamefont{Giovannetti et~al.}(2004)\citenamefont{Giovannetti,
  Lloyd, and Maccone}}]{giovannetti2004quantum}
\bibinfo{author}{\bibfnamefont{V.}~\bibnamefont{Giovannetti}},
  \bibinfo{author}{\bibfnamefont{S.}~\bibnamefont{Lloyd}}, \bibnamefont{and}
  \bibinfo{author}{\bibfnamefont{L.}~\bibnamefont{Maccone}},
  \bibinfo{journal}{Science} \textbf{\bibinfo{volume}{306}},
  \bibinfo{pages}{1330} (\bibinfo{year}{2004}).

\bibitem[{\citenamefont{Pezz{\'e} and Smerzi}(2009)}]{pezze2009entanglement}
\bibinfo{author}{\bibfnamefont{L.}~\bibnamefont{Pezz{\'e}}} \bibnamefont{and}
  \bibinfo{author}{\bibfnamefont{A.}~\bibnamefont{Smerzi}},
  \bibinfo{journal}{Phys. Rev. Lett.} \textbf{\bibinfo{volume}{102}},
  \bibinfo{pages}{100401} (\bibinfo{year}{2009}).

\bibitem[{\citenamefont{Itano~{\it{et al}}}(1993)}]{itano1993quantum}
\bibinfo{author}{\bibfnamefont{W.}~\bibnamefont{Itano~{\it{et al}}}},
  \bibinfo{journal}{Phys. Rev. A} \textbf{\bibinfo{volume}{47}},
  \bibinfo{pages}{3554} (\bibinfo{year}{1993}).

\bibitem[{\citenamefont{Giovannetti et~al.}(2006)\citenamefont{Giovannetti,
  Lloyd, and Maccone}}]{giovannetti2006quantumprlversion}
\bibinfo{author}{\bibfnamefont{V.}~\bibnamefont{Giovannetti}},
  \bibinfo{author}{\bibfnamefont{S.}~\bibnamefont{Lloyd}}, \bibnamefont{and}
  \bibinfo{author}{\bibfnamefont{L.}~\bibnamefont{Maccone}},
  \bibinfo{journal}{Phys. Rev. Lett.} \textbf{\bibinfo{volume}{96}},
  \bibinfo{pages}{10401} (\bibinfo{year}{2006}).

\bibitem[{\citenamefont{Nagata~{\it{et al}}}(2007)}]{nagata2007beatingsimple}
\bibinfo{author}{\bibfnamefont{T.}~\bibnamefont{Nagata~{\it{et al}}}},
  \bibinfo{journal}{Science} \textbf{\bibinfo{volume}{316}},
  \bibinfo{pages}{726} (\bibinfo{year}{2007}).

\bibitem[{\citenamefont{Afek et~al.}(2010)\citenamefont{Afek, Ambar, and
  Silberberg}}]{afek2010high}
\bibinfo{author}{\bibfnamefont{I.}~\bibnamefont{Afek}},
  \bibinfo{author}{\bibfnamefont{O.}~\bibnamefont{Ambar}}, \bibnamefont{and}
  \bibinfo{author}{\bibfnamefont{Y.}~\bibnamefont{Silberberg}},
  \bibinfo{journal}{Science} \textbf{\bibinfo{volume}{328}},
  \bibinfo{pages}{879} (\bibinfo{year}{2010}).

\bibitem[{\citenamefont{Kok et~al.}(2002)\citenamefont{Kok, Lee, and
  Dowling}}]{kok2002creation}
\bibinfo{author}{\bibfnamefont{P.}~\bibnamefont{Kok}},
  \bibinfo{author}{\bibfnamefont{H.}~\bibnamefont{Lee}}, \bibnamefont{and}
  \bibinfo{author}{\bibfnamefont{J.}~\bibnamefont{Dowling}},
  \bibinfo{journal}{Phys. Rev. A} \textbf{\bibinfo{volume}{65}},
  \bibinfo{pages}{52104} (\bibinfo{year}{2002}).

\bibitem[{\citenamefont{Leibfried~{\it{et
  al}}}(2005)}]{leibfried2005creationsimple}
\bibinfo{author}{\bibfnamefont{D.}~\bibnamefont{Leibfried~{\it{et al}}}},
  \bibinfo{journal}{Nature} \textbf{\bibinfo{volume}{438}},
  \bibinfo{pages}{639} (\bibinfo{year}{2005}).

\bibitem[{\citenamefont{Jones~{\it{et al}}}(2009)}]{jones2009magneticsimple}
\bibinfo{author}{\bibfnamefont{J.}~\bibnamefont{Jones~{\it{et al}}}},
  \bibinfo{journal}{science} \textbf{\bibinfo{volume}{324}},
  \bibinfo{pages}{1166} (\bibinfo{year}{2009}).

\bibitem[{\citenamefont{Leibfried et~al.}(2004)\citenamefont{Leibfried,
  Barrett, Schaetz, Britton, Chiaverini, Itano, Jost, Langer, and
  Wineland}}]{leibfried2004toward}
\bibinfo{author}{\bibfnamefont{D.}~\bibnamefont{Leibfried}},
  \bibinfo{author}{\bibfnamefont{M.}~\bibnamefont{Barrett}},
  \bibinfo{author}{\bibfnamefont{T.}~\bibnamefont{Schaetz}},
  \bibinfo{author}{\bibfnamefont{J.}~\bibnamefont{Britton}},
  \bibinfo{author}{\bibfnamefont{J.}~\bibnamefont{Chiaverini}},
  \bibinfo{author}{\bibfnamefont{W.}~\bibnamefont{Itano}},
  \bibinfo{author}{\bibfnamefont{J.}~\bibnamefont{Jost}},
  \bibinfo{author}{\bibfnamefont{C.}~\bibnamefont{Langer}}, \bibnamefont{and}
  \bibinfo{author}{\bibfnamefont{D.}~\bibnamefont{Wineland}},
  \bibinfo{journal}{Science} \textbf{\bibinfo{volume}{304}},
  \bibinfo{pages}{1476} (\bibinfo{year}{2004}).

\bibitem[{\citenamefont{Schaffry et~al.}(2010)\citenamefont{Schaffry, Gauger,
  Morton, Fitzsimons, Benjamin, and
  Lovett}}]{schaffry2010ensemblebasedmetrology}
\bibinfo{author}{\bibfnamefont{M.}~\bibnamefont{Schaffry}},
  \bibinfo{author}{\bibfnamefont{E.}~\bibnamefont{Gauger}},
  \bibinfo{author}{\bibfnamefont{J.}~\bibnamefont{Morton}},
  \bibinfo{author}{\bibfnamefont{J.}~\bibnamefont{Fitzsimons}},
  \bibinfo{author}{\bibfnamefont{S.}~\bibnamefont{Benjamin}}, \bibnamefont{and}
  \bibinfo{author}{\bibfnamefont{B.}~\bibnamefont{Lovett}},
  \bibinfo{journal}{arXiv:1007.2491}  (\bibinfo{year}{2010}).

\bibitem[{\citenamefont{Simmons et~al.}(2009)\citenamefont{Simmons, Jones,
  Karlen, Ardavan, and Morton}}]{simmons2009magnetic0907}
\bibinfo{author}{\bibfnamefont{S.}~\bibnamefont{Simmons}},
  \bibinfo{author}{\bibfnamefont{J.}~\bibnamefont{Jones}},
  \bibinfo{author}{\bibfnamefont{S.}~\bibnamefont{Karlen}},
  \bibinfo{author}{\bibfnamefont{A.}~\bibnamefont{Ardavan}}, \bibnamefont{and}
  \bibinfo{author}{\bibfnamefont{J.}~\bibnamefont{Morton}},
  \bibinfo{journal}{arXiv:0907.1372}  (\bibinfo{year}{2009}).

\bibitem[{\citenamefont{Dur and Briegel}(2004)}]{dur2004stabilitymacro}
\bibinfo{author}{\bibfnamefont{W.}~\bibnamefont{Dur}} \bibnamefont{and}
  \bibinfo{author}{\bibfnamefont{H.}~\bibnamefont{Briegel}},
  \bibinfo{journal}{Phys. Rev. Lett.} \textbf{\bibinfo{volume}{92}},
  \bibinfo{pages}{180403} (\bibinfo{year}{2004}).

\bibitem[{\citenamefont{Shaji and Caves}(2007)}]{shajicaves2007qubit}
\bibinfo{author}{\bibfnamefont{A.}~\bibnamefont{Shaji}} \bibnamefont{and}
  \bibinfo{author}{\bibfnamefont{C.}~\bibnamefont{Caves}},
  \bibinfo{journal}{Phys. Rev. A} \textbf{\bibinfo{volume}{76}},
  \bibinfo{pages}{32111} (\bibinfo{year}{2007}).

\bibitem[{\citenamefont{Huelga et~al.}(1997)\citenamefont{Huelga, Macchiavello,
  Pellizzari, Ekert, Plenio, and Cirac}}]{huelga1997improvement}
\bibinfo{author}{\bibfnamefont{S.}~\bibnamefont{Huelga}},
  \bibinfo{author}{\bibfnamefont{C.}~\bibnamefont{Macchiavello}},
  \bibinfo{author}{\bibfnamefont{T.}~\bibnamefont{Pellizzari}},
  \bibinfo{author}{\bibfnamefont{A.}~\bibnamefont{Ekert}},
  \bibinfo{author}{\bibfnamefont{M.}~\bibnamefont{Plenio}}, \bibnamefont{and}
  \bibinfo{author}{\bibfnamefont{J.}~\bibnamefont{Cirac}},
  \bibinfo{journal}{Phys. Rev. Lett.} \textbf{\bibinfo{volume}{79}},
  \bibinfo{pages}{3865} (\bibinfo{year}{1997}).

\bibitem[{\citenamefont{Gardiner and Zoller}(2004)}]{GZ01b}
\bibinfo{author}{\bibfnamefont{C.~W.} \bibnamefont{Gardiner}} \bibnamefont{and}
  \bibinfo{author}{\bibfnamefont{P.}~\bibnamefont{Zoller}},
  \emph{\bibinfo{title}{Quantum Noise}} (\bibinfo{publisher}{Springer, Berlin},
  \bibinfo{year}{2004}).

\bibitem[{\citenamefont{Nakazato et~al.}(1996)\citenamefont{Nakazato, Namiki,
  and Pascazio}}]{NakazatoNamikiPascazio01a}
\bibinfo{author}{\bibfnamefont{H.}~\bibnamefont{Nakazato}},
  \bibinfo{author}{\bibfnamefont{M.}~\bibnamefont{Namiki}}, \bibnamefont{and}
  \bibinfo{author}{\bibfnamefont{S.}~\bibnamefont{Pascazio}},
  \bibinfo{journal}{Int. J. Mod. B} \textbf{\bibinfo{volume}{10}},
  \bibinfo{pages}{247} (\bibinfo{year}{1996}).

\bibitem[{\citenamefont{Schulman}(1997)}]{Schulman01a}
\bibinfo{author}{\bibfnamefont{L.~S.} \bibnamefont{Schulman}},
  \bibinfo{journal}{J. Phys. A} \textbf{\bibinfo{volume}{30}},
  \bibinfo{pages}{L293} (\bibinfo{year}{1997}).

\bibitem[{\citenamefont{Meeron}(1957)}]{Meeron01a}
\bibinfo{author}{\bibfnamefont{E.}~\bibnamefont{Meeron}}, \bibinfo{journal}{J.
  Chem. Phys.} \textbf{\bibinfo{volume}{27}}, \bibinfo{pages}{67}
  (\bibinfo{year}{1957}).

\bibitem[{\citenamefont{Yoshihara et~al.}(2006)\citenamefont{Yoshihara,
  Harrabi, Niskanen, and Nakamura}}]{YoshiharaHarrabiNiskanenNakamura01a}
\bibinfo{author}{\bibfnamefont{F.}~\bibnamefont{Yoshihara}},
  \bibinfo{author}{\bibfnamefont{K.}~\bibnamefont{Harrabi}},
  \bibinfo{author}{\bibfnamefont{A.}~\bibnamefont{Niskanen}}, \bibnamefont{and}
  \bibinfo{author}{\bibfnamefont{Y.}~\bibnamefont{Nakamura}},
  \bibinfo{journal}{Phys. Rev. Lett.} \textbf{\bibinfo{volume}{97}},
  \bibinfo{pages}{167001} (\bibinfo{year}{2006}).

\bibitem[{\citenamefont{Kakuyanagi et~al.}(2007)\citenamefont{Kakuyanagi, Meno,
  Saito, Nakano, Semba, Takayanagi, Deppe, and
  Shnirman}}]{KakuyanagiMenoSaitoNakanoSembaTakayanagiDeppeShnirman01a}
\bibinfo{author}{\bibfnamefont{K.}~\bibnamefont{Kakuyanagi}},
  \bibinfo{author}{\bibfnamefont{T.}~\bibnamefont{Meno}},
  \bibinfo{author}{\bibfnamefont{S.}~\bibnamefont{Saito}},
  \bibinfo{author}{\bibfnamefont{H.}~\bibnamefont{Nakano}},
  \bibinfo{author}{\bibfnamefont{K.}~\bibnamefont{Semba}},
  \bibinfo{author}{\bibfnamefont{H.}~\bibnamefont{Takayanagi}},
  \bibinfo{author}{\bibfnamefont{F.}~\bibnamefont{Deppe}}, \bibnamefont{and}
  \bibinfo{author}{\bibfnamefont{A.}~\bibnamefont{Shnirman}},
  \bibinfo{journal}{Phys. Rev. Lett.} \textbf{\bibinfo{volume}{98}},
  \bibinfo{pages}{047004} (\bibinfo{year}{2007}).

\bibitem[{\citenamefont{Matsuzaki et~al.}(2010)\citenamefont{Matsuzaki, Saito,
  Kakuyanagi, and Semba}}]{matsuzaki2010quantumzenontt}
\bibinfo{author}{\bibfnamefont{Y.}~\bibnamefont{Matsuzaki}},
  \bibinfo{author}{\bibfnamefont{S.}~\bibnamefont{Saito}},
  \bibinfo{author}{\bibfnamefont{K.}~\bibnamefont{Kakuyanagi}},
  \bibnamefont{and} \bibinfo{author}{\bibfnamefont{K.}~\bibnamefont{Semba}},
  \bibinfo{journal}{Phys. Rev. B} \textbf{\bibinfo{volume}{82}},
  \bibinfo{pages}{180518} (\bibinfo{year}{2010}).

\bibitem[{\citenamefont{Palma et~al.}(1996)\citenamefont{Palma, Suominen, and
  Ekert}}]{PSE}
\bibinfo{author}{\bibfnamefont{G.~M.} \bibnamefont{Palma}},
  \bibinfo{author}{\bibfnamefont{K.~A.} \bibnamefont{Suominen}},
  \bibnamefont{and} \bibinfo{author}{\bibfnamefont{A.~K.} \bibnamefont{Ekert}},
  \bibinfo{journal}{Proc. R. Soc. London. Ser.A}
  \textbf{\bibinfo{volume}{452}}, \bibinfo{pages}{567} (\bibinfo{year}{1996}).

\bibitem[{\citenamefont{Duan and Guo}(1997)}]{duan1997perturbativefidelity}
\bibinfo{author}{\bibfnamefont{L.}~\bibnamefont{Duan}} \bibnamefont{and}
  \bibinfo{author}{\bibfnamefont{G.}~\bibnamefont{Guo}},
  \bibinfo{journal}{Phys. Rev. A} \textbf{\bibinfo{volume}{56}},
  \bibinfo{pages}{4466} (\bibinfo{year}{1997}).

\bibitem[{\citenamefont{Kane}(1998)}]{K01a}
\bibinfo{author}{\bibfnamefont{B.~E.} \bibnamefont{Kane}},
  \bibinfo{journal}{Nature} \textbf{\bibinfo{volume}{393}},
  \bibinfo{pages}{131} (\bibinfo{year}{1998}).

\bibitem[{\citenamefont{Ladd et~al.}(2005)\citenamefont{Ladd, Maryenko,
  Yamamoto, Abe, and Itoh}}]{ladd2005coherence}
\bibinfo{author}{\bibfnamefont{T.}~\bibnamefont{Ladd}},
  \bibinfo{author}{\bibfnamefont{D.}~\bibnamefont{Maryenko}},
  \bibinfo{author}{\bibfnamefont{Y.}~\bibnamefont{Yamamoto}},
  \bibinfo{author}{\bibfnamefont{E.}~\bibnamefont{Abe}}, \bibnamefont{and}
  \bibinfo{author}{\bibfnamefont{K.}~\bibnamefont{Itoh}},
  \bibinfo{journal}{Phys. Rev. B} \textbf{\bibinfo{volume}{71}},
  \bibinfo{pages}{14401} (\bibinfo{year}{2005}).

\end{thebibliography}

\end{document}